\documentclass[twocolumn,showpacs,preprintnumbers,amsmath,amssymb]{revtex4}

\usepackage{graphicx}% Include figure files
\usepackage{dcolumn}% Align table columns on decimal point
\usepackage{bm}% bold math

\begin{document}
\widetext

\title{Thermodynamic phase diagram of Fe(Se$_{0.5}$Te$_{0.5}$) single crystals up to 28 Tesla}

\author{T. Klein$^{1,2}$, D. Braithwaite$^3$, A. Demuer$^4$, W. Knafo$^5$,  G. Lapertot$^3$, C. Marcenat$^3$, P. Rodi\`ere$^1$, and I. Sheikin$^4$, P. Strobel$^1$, A. Sulpice$^1$ and P. Toulemonde$^1$}

\address{$^{1}$ Institut N\'eel, CNRS, 25 rue des martyrs, 38042 
Grenoble, France}
\address{$^{2}$ Institut Universitaire de France and Universit\'e J.Fourier-Grenoble 1, France}
\address{$^3$ SPSMS, UMR-E9001, CEA-INAC/ UJF-Grenoble 1, 17 rue des 
martyrs, 38054 Grenoble, France} 
\address{$^{4}$ LNCMI-CNRS, 25 Avenue des Martyrs, BP 166, 38042 Grenoble, France}
\address{$^5$ LNCMI, UPR 3228, CNRS-UJF-UPS-INSA, 31400 Toulouse, France}
\date{\today}

\begin{abstract}
We report on specific heat ($C_p$), transport, Hall probe and penetration depth measurements performed on Fe(Se$_{0.5}$Te$_{0.5}$) single crystals ($T_c \sim 14$ K). The thermodynamic upper critical field $H_{c2}$ lines has been deduced from $C_p$ measurements up to 28 T for both $H\|c$ and $H\|ab$, and compared to the lines deduced from transport measurements (up to 55 T in pulsed magnetic fields). We show that this {\it  thermodynamic} $H_{c2}$ line presents a very strong downward curvature for $T \rightarrow T_c$ which is not visible in transport measurements. This temperature dependence associated to an upward curvature of the field dependence of the Sommerfeld coefficient  confirm that $H_{c2}$ is limited by paramagnetic effects. Surprisingly this paramagnetic limit is visible here up to $T/T_c \sim 0.99$  (for $H\|ab$) which is the consequence of a very small value of the coherence length $\xi_c(0) \sim 4 \AA$ (and $\xi_{ab}(0) \sim 15 \AA$),  confirming the strong renormalisation of the effective mass (as compared to DMFT calculations) previously observed in ARPES measurements [Phys. Rev. Lett. 104, 097002 (2010)]. $H_{c1}$ measurements lead to $\lambda_{ab}(0) = 430 \pm 50$ nm and $\lambda_c(0) = 1600 \pm 200$ nm and the corresponding anisotropy is approximatively temperature independent ($\sim 4$), being close to the anisotropy of $H_{c2}$ for $T\rightarrow T_c$. The temperature dependence of both $\lambda$ ($\propto T^2$) and the electronic contribution to the specific heat confirm the non conventional coupling mechanism in this system.
\end{abstract}

\maketitle

\section{introduction}
The discovery of superconductivity up to 55K in iron-based systems \cite{Kamihara} has generated
tremendous interest. Among those, iron selenium (FeSe$_{1-\delta}$) \cite{Hsu} has been reported to be superconducting with a critical temperature of 8 K at ambient pressure, rising to 34-37 K under 7-15 GPa \cite{Garbarino}. On the other hand, the substitution of tellurium on the selenium site in Fe$_{1+\delta}$(Te$_x$Se$_{1-x}$) increases $T_c$ to a maximum on the order of 14-15 K at ambient pressure \cite{Sales,Yeh} (for $x \sim 0.5$). This binary compound is very interesting as it shares the most salient characteristics of iron based systems (square-planar lattice of Fe with tetrahedral coordination)  but has the simplest crystallographic structure among Fe-based superconductors (no charge reservoir \cite{Fe2}, so-called 11-structure). Moreover, even though the endpoint Fe$_{1+\delta}$Te \cite{Bao} compound displays antiferromagnetic ordering,  a magnetic resonance similar to that observed in other parent compounds (with a (1/2, 1/2) nesting vector connecting the $\Gamma$ and $M$ points of the Fermi surface) is recovered for intermediate Te contents \cite{Qiu,Martinelli} suggesting a common mechanism for superconductivity in all iron based superconductors. However, in contrast to iron pnictides which show only weak to moderate correlations, recent ARPES measurements suggested the existence of very large mass renormaliszation factors (up to $\sim 20$ as compared to DMFT calculations) \cite{Tamai} indicating that Fe(Se,Te) is a strongly correlated metal differing significantly from iron pnictides.

\begin{figure} 
\begin{center}
\resizebox{0.48\textwidth}{!}{\includegraphics{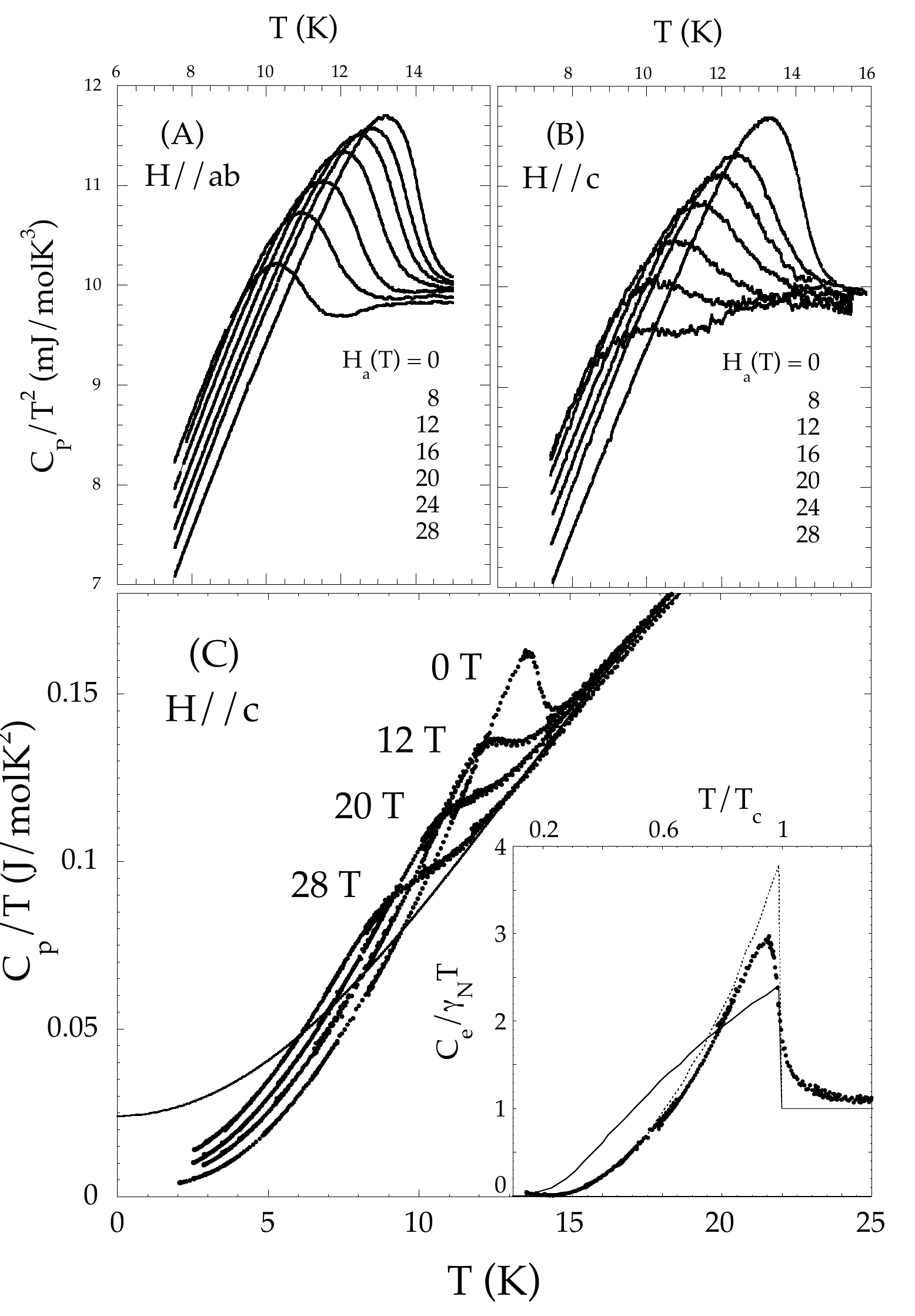}}
\caption{AC specific heat measurements $C_p/T^2$ as a function of $T$ of a Fe(Se$_{0.5}$Te$_{0.5}$) single crystal (sample A4) for $\mu_0H= 0$, $8$, $12$, $16$, $20$, $24$ and $28$ T (from right to left) for $H\|ab$ (A) and $H\|c$ (B). The data have been renormalized taking $C_p(T=20K)= 3.8$ J/molK. The $H_{c2}$ line is deduced from the midpoint of the specific heat jump after subtraction of a smooth polynomial background. (C) : Specific heat from relaxation data ($H\|c$) for the indicated magnetic fields (sample A5).  Inset : temperature dependence of the electronic contribution to the specific heat $Ce = C_p-\beta T^3 - \delta T^5$ (solid symbols) where the phonon contribution ($\beta T^3 + \delta T^5$) has been subtracted from the normal state data (see thin line in Fig.1C).  The BCS behavior for $2\delta/kT_c = 3.5$ (solid line) and $2\Delta/kT_c=5$ (dotted line) are  displayed for comparison. }
\label{Fig.1}
\end{center}
\end{figure}

In order to shed light on superconductivity in these systems, it is of fundamental importance to obtain a precise determination of both upper and lower critical fields and their anisotropy. Up to now $H_{c2}$ has mainly been deduced from transport measurements \cite{Lei,Fang,Braithwaite}, and more recently by specific heat up to 14 T \cite{Sarafin}. As in other pnictides (see \cite{Pribulova} and references therein), high $H_{c2}(0)$ values have been reported but, in the case of Fe(Te$_x$Se$_{1-x}$), strong deviations from the standard Werthamer-Helfand-Hohenberg model for $H_{c2}(T)$ have been reported. Those deviations have been associated to paramagnetic limitations (so-called Pauli limit) \cite{Lei,Fang,Braithwaite}. However, in presence of strong thermal fluctuations (see discussion below), the determination of $H_{c2}$ from transport measurement becomes very hazardous and a thorough analysis was hence lacking of an unambiguous determination of $H_{c2}$ from specific heat measurements. 

 We show that the $H_{c2}$ lines actually display a very strong downwards curvature close to $T_c$ corresponding to $\mu_0dH_{c2}/dT$ values rising up to $\sim 12$ T/K for $H\|c$ and even $\sim 45$ T/K for $H\|ab$. This strong curvature, not visible in transport data, shows that $H_{c2}$ remains limited by paramagnetic effects up to temperatures very close to $T_c$ (up to $T/T_c \sim 0.99$ for $H\|ab$). The corresponding Pauli field $H_p$ is slightly anisotropic ($H_p^{\|ab}/H_p^{\|c} \sim 0.8$) whereas the orbital limit ($H_o(0)$) presents a much stronger anisotropy $H_o(0)^{\|ab}/H_o(0)^{\|c} \sim 3-4$. The huge $\mu_0H_o(0)$ values ($\sim 130 \pm 20$ T for $H\|c$ and $\sim 400 \pm 50$ T for $H\|ab$) correspond to very small coherence length values ($\xi_{ab}(0) \sim 15 \pm 1 \AA$ and $\xi_c(0) \sim 4 \pm 1 \AA$) confirming the large value of the effective mass previously observed by ARPES \cite{Tamai} and hence supporting the presence of strong electronic correlations in this system. 
 
 In addition, preliminary $H_{c1}$ measurements led to contradictory results. On the one hand, Yadav {\it et al.} \cite{Yadav} reported on rather high $H_{c1}$ values $\sim 100$ G and $\sim 400$ G for $H\|c$ and $H\|ab$ respectively with $H_{c1}$ lines showing a clear upward curvature at low temperature. On the other hand, Bendele {\it et al.} \cite{Bendele} obtained much smaller values ($\sim 20$ G and $\sim 45$ G for $H\|c$ and $H\|ab$ respectively) associated with a clear saturation  of the $H_{c1}(T)$ lines at low temperature. Finally, Kim {\it et al.} \cite{Kim} reported on strong deviations of the temperature dependance of the superfluid density ($\rho_s \propto 1/\lambda^2 \propto H_{c1}$) from the standard behavior, attributed to a clear signature of multigap superconductivity. We present here detailed first penetration field measurements performed with Hall sensor arrays in a variety of single crystals showing very different aspect ratios. We hence obtained $\mu_0H_{c1}^{\|c}(0) = 78 \pm 5$ G and $\mu_0H_{c1}^{\|ab}(0) = 23 \pm 3$ G. The $H_{c1}$ lines clearly flatten off at low temperature but do not show the pronounced deep previously obtained in Tunnel Diode Oscillator (TDO) measurements \cite{Kim}. Our TDO measurements however led to a similar deep which is probably due to an overestimation of the absolute $\Delta\lambda(T)$  value related to spurious edge effects. Finally, we obtained a temperature independent $\Gamma_{H_{c1}}= H_{c1}^{\|c}/ H_{c1}^{\|ab}$ values $\sim 3.3 \pm 0.5$ which corresponds to $\Gamma_\lambda=\lambda_{c}/\lambda_{ab} \sim 4.0 \pm 0.8$ (see below), being close to the $\gamma_{H_{c2}}$ value obtained for $T \rightarrow T_c$ (i.e. $\sim H_o^{\|ab}/H_o^{\|c}$). 
 
Finally, we confirm that $\lambda \propto T^2$ in both crystallographic directions and show that the temperature dependence of $C_p$ strongly deviates from the standard BCS weak coupling behavior confirming the non conventional coupling mechanism of this system. However, the amplitude of the specific heat jump is much larger than those previously reported in other Fe(Se,Te) samples and hence does not follow the $\Delta C_p$ vs $T_c^3$ scaling law reported in iron based systems  \cite{Budko,Paglione}.

\section{Sample preparation and experiments}

We present here specific heat, transport, Hall probe and Tunnel Diode Oscillator (penetration depth) measurements performed in Fe$_{1+\delta}$(Se$_{0.50}$Te$_{0.5}$) single crystals grown by two different techniques. Samples A have been grown using the sealed quartz tube method. The samples were prepared 
from very pure iron and tellurium pieces and selenium shots in a 1:0.5:0.5 ratio, loaded together in a quartz tube which has been sealed
under vacuum. The elements were heated slowly (100\r{ }C/h) at 500\r{ }C for 10 h, then melted at 1000\r{ }C for 20h, cooled slowly down to 350\r{ }C at 5\r{ }C/h, and finally cooled faster by switching off the furnace. Single crystals were extracted
mechanically from the resulting ball, the crystals being easy cleaved perpendicular to their c crystallographic axis. The refined lattice parameters of the Fe$_{1+\delta}$(Se$_{0.5}$Te$_{0.5}$) tetragonal main phase, a = 3.7992(7) ~{\AA} and c = 6.033(2) ~{\AA}, are in agreement with the literature~\cite{Sales,Bendele}. The real composition of the crystals checked by x-ray energy dispersive micro-analysis using a scanning electron microscope was found to be Fe$_{1.05(2)}$(Te$_{0.55(2)}$Se$_{0.45(2)}$). The temperature dependence of the resistivity shows a metallic behavior at low temperature as expected for this low level ($\delta$ = 0.05) of interstitial iron \cite{Liu}.

Samples of batch B were grown with the Bridgman technique using a double wall quartz ampoule. The inside tube had a tipped bottom with a 30$^\circ$ angle and an open top. The inside wall of the outer ampoule was carbon coated to achieve the lowest possible oxygen partial pressure during the growth. The Bridgman ampoule was inserted in a three zone gradient furnace (1000$^\circ$/840$^\circ$ /700$^\circ$) and lowered at a speed of 3 mm/h. At the end of the growth, temperature was lowered to room temperature at 50 $^\circ$ C/h. Further characterizations of these crystals have been published elsewhere \cite{Braithwaite}. The different single crystals used in this study have been listed in Table 1.

\begin{table}
\caption{\label{table1}  average thickness $d$, width $w$ length $l$ or mass $m$ of the samples and measurement techniques  [C$_p$ = specific heat, R = transport, HP = Hall probe and TDO = tunnel diode oscillator]}
\begin{ruledtabular}
\begin{tabular}{ccccc}
Sample&$d$ ($\mu$m)&$w$ ($\mu$m)&$l$ ($\mu$m)&measured by\\
\hline
A1 & 50 & 180 & 220 & AC-C$_p$, HP, TDO\\
A2 & 60 & 300 & 750 & HP, TDO \\
A3 & 65 & 400 & 600 & HP, TDO \\
A3' & 40 & 400 & 300 & HP, TDO \\
A3'' & 40 & 100 & 100 & HP, TDO \\
\hline
A4&\multicolumn{3}{c}{$m\sim 50$ $\mu$g}&AC-C$_p$\\
A5&\multicolumn{3}{c}{$m\sim 1.1$ mg}&DC-C$_p$, R\\
B1&\multicolumn{3}{c}{$m\sim 0.7 $ mg}&DC-C$_p$, R\\
\end{tabular}
\end{ruledtabular}
\end{table}

\begin{figure}
\begin{center}
\resizebox{0.48\textwidth}{!}{\includegraphics{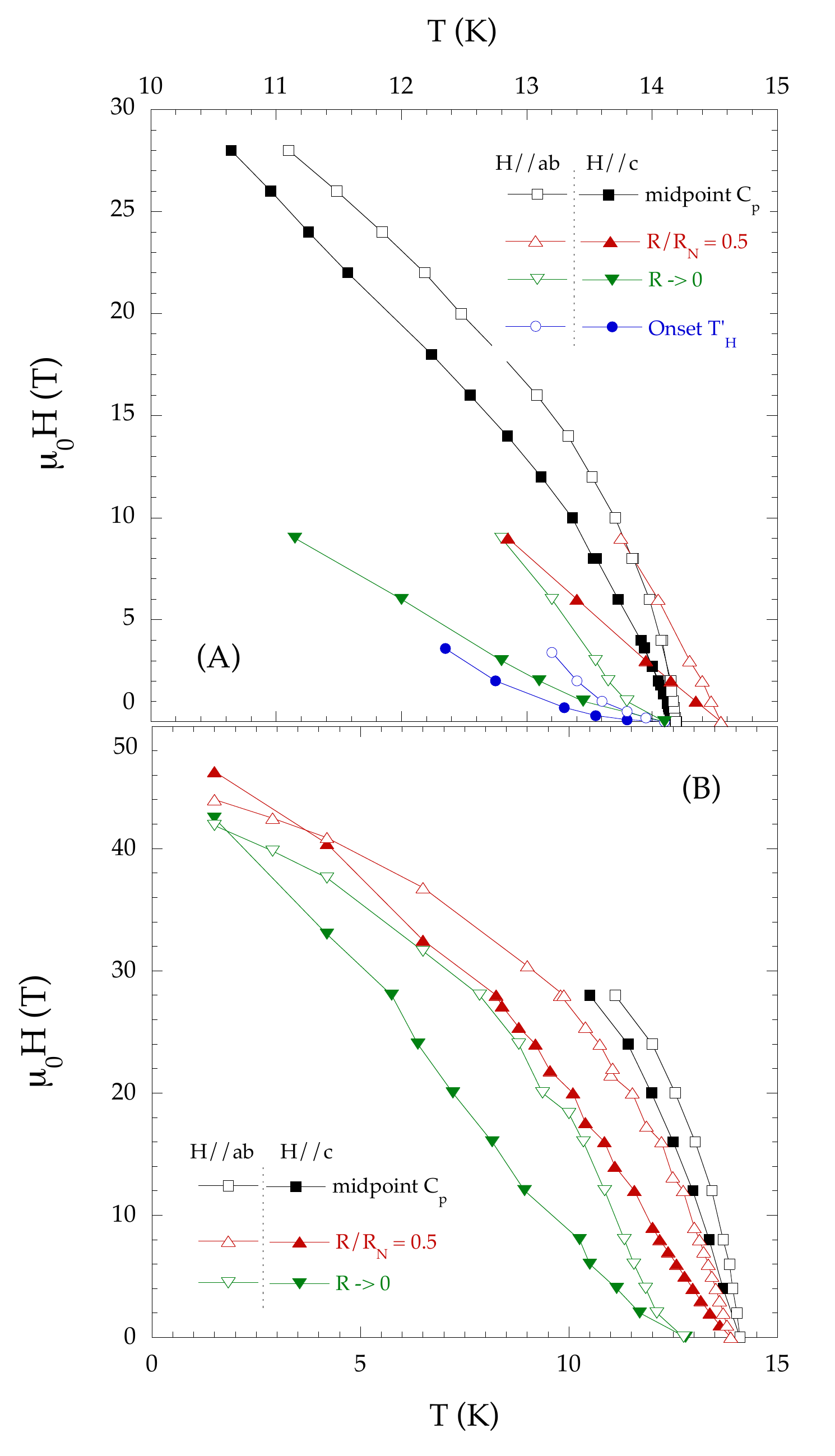}}
\caption{(color on line) (A) $H-T$ phase diagram (batch A) in both $H\|ab$  (open symbols) and $H\|c$ (closed symbols) displaying the $H_{c2}$ line deduced from specific heat measurements (see Fig.1)(squares), the field corresponding to the onset of the diamagnetic screening (circles) as well as the fields corresponding to zero resistance (downward triangles) and $R/R_N=0.5$ (upward triangle, $R_N$ being the normal state resistance). (B) same as in (A) for sample B1. The transport data are taken from [13]. See Fig.3 for a direct comparison between the $H_{c2}$ lines in each batch.}
\label{Fig.2}
\end{center}
\end{figure}

The $C_p$ measurements have been performed in magnetic fields up to 28 T using both an AC high 
sensitivity technique and a conventional relaxation technique. For AC measurements, heat was supplied to the sample by a light emitting diode via an optical fiber and the corresponding temperature oscillations were recorded with a thermocouple (sample A1 and A4). In parallel, the specific heat (sample A5 and B1) were carried out in a miniaturized high-resolution micro-calorimeter using the ÔÔlong-relaxation
techniqueÕÕ. The chip resistance used as both thermometer and heater as well as the
thermal conductance of its leads have been carefully calibrated up to 28T
using a capacitance thermometer. Each relaxation provides about 1000 data points over a temperature
interval of about $80\%$ above the base temperature which has been varied
between 1.8 and 20 K. Data can be recorded during heating and cooling.
The merging of the upward and downward relaxation data provides a highly
reliable check of the accuracy of this method.

Electrical transport measurements have been performed on sample B1 in static magnetic fields up to 28T and pulsed magnetic fields up to 55T and are described in detail elsewhere \cite{Braithwaite}. We have also measured the resistivity of sample A5 with a commercial device (PPMS) up to 9T.

The real part of the AC transmittivity, $T_{H}^{\prime}$, of samples A1 to A3'' has been measured 
by centering these on a miniature GaAs-based quantum well Hall Sensor 
(of dimension $8 \times 8$ $\mu$m$^2$). The sensor is used to record the time-varying 
component $B_{ac}$ of the local magnetic induction as the sample is exposed to an ac field $ \sim 1$ Oe ( $\omega \sim 210$ Hz). $T'_H$ is then defined as : $T^{\prime}_{H} = [B_{ac}(T) - B_{ac}(4.2 {\mathrm  K})]/[ B_{ac}(T \gg T_{c})  - B_{ac}(4.2 {\mathrm K})]$. The remanent local DC field ($B_{rem}(H_a)$) in the sample has been measured after applying a magnetic field $H_a$ and sweeping the field back to zero. In the Meissner state, no vortices penetrate the sample and $B_{rem}$ remains equal to zero up to $H_a$ = $H_f$ (the first penetration field). A finite remanent field is then obtained for field amplitudes larger than $H_f$ as vortices remain pinned in the sample. 

Finally, the London magnetic penetration depth  in the Meissner state , $\lambda$, has been measured on the same samples with a LC oscillating circuit (14MHz) driven by a Tunnel Diode (TDO). The samples have been glued at the bottom of a sapphire rod which were introduced in a coil of inductance $L$. The variation of the penetration depth induce a change in $L$ and hence a shift of the resonant frequency $\delta f(T)=f(T)-f(T_{min})$. $\delta f(T)$, renormalised to the frequency shift corresponding to the extraction of the sample from the coil $\Delta f_0$ is then equal to the magnetic susceptibility. At low temperatures (typically for T $\leq 12K$), $\lambda<<d$ (d being the lowest dimension of the sample, here the thickness), and we have $\frac{\delta f(T)}{\Delta f_0}=\frac{\tilde{\lambda}}{\tilde{R}}$ where $\tilde{\lambda}$ is an effective penetration depth depending on the field orientation and $\tilde{R}$ an effective dimension of the sample. When the magnetic field is applied along the c-axis, only the in-plane supercurrents are probed and $\tilde{\lambda}=\lambda_{ab}$, whereas $\tilde{\lambda}=\lambda_{ab}+\frac{d}{w}\lambda_{c}$ for H//ab ($w$ being the width of the sample). The effective dimension $\tilde{R}$ is calculated following \cite{Prozorov}. 

 \begin{figure}
\begin{center}
\resizebox{0.46\textwidth}{!}{\includegraphics{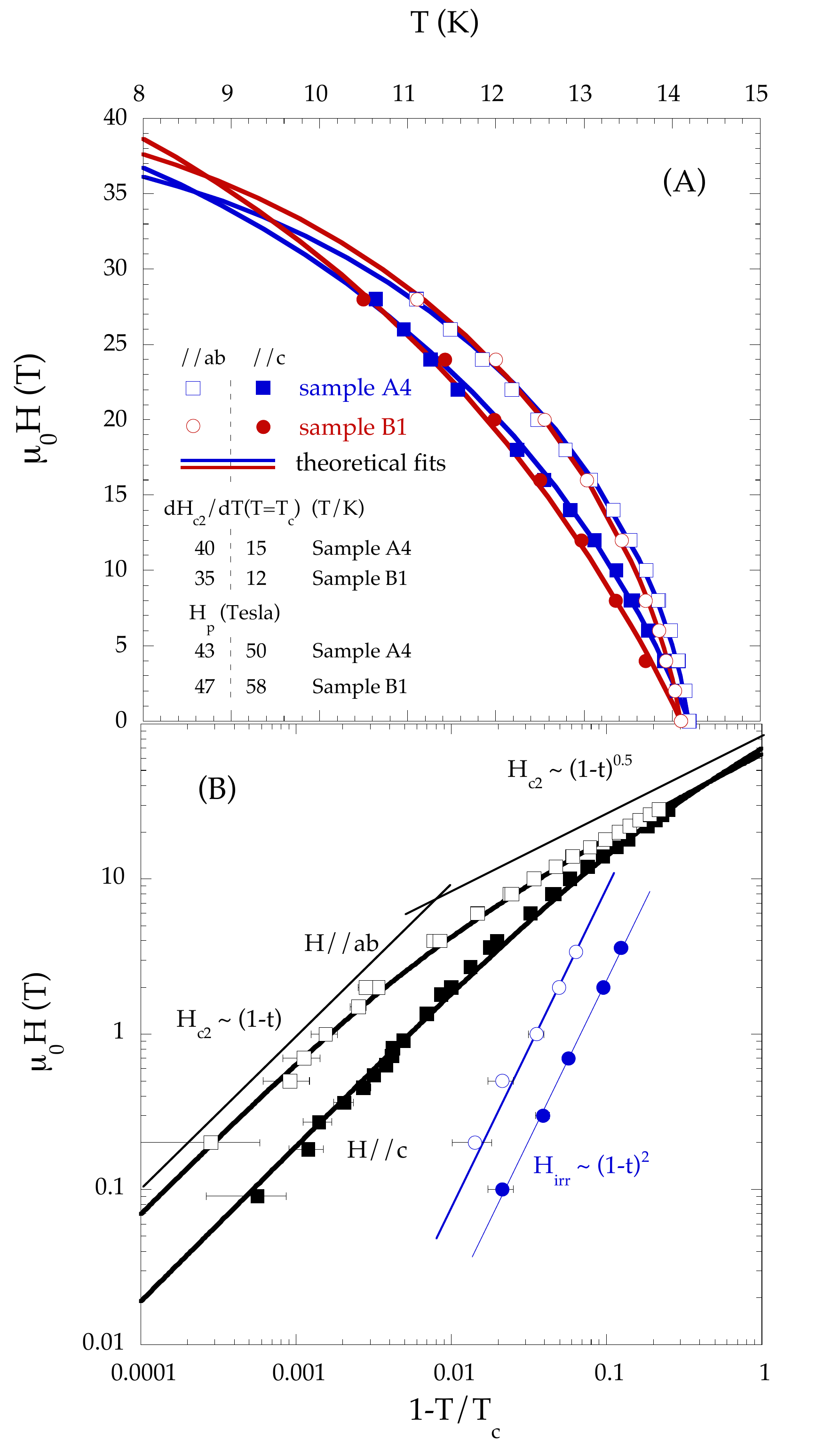}}
\caption{(color online) (A) comparison between the $H_{c2}(T)$ values deduced from specific heat measurements in sample A4 (blue squares) and sample B1 (red circles) for both $H\|c$ (closed symbols) and $H\|ab$ (open symbols) and theoretical values for  clean weakly coupled BCS superconductors including both orbital and Pauli limitations (solid lines). (B)  $H_{c2}$ vs $(1-t)$ (sample A4) in a log-log scale ($t=T/T_c$) showing that the linear dependence of the $H_{c2}$ line rapidly crosses to a $\sim (1-t)^{0.5}$ dependence. The solid lines are fits to Eq.(1) (see text for details). The linear slopes close to $T_c$ ($\sim 45$ T/K and $\sim 12$ T/K) extrapolate to very high orbital limits. On the contrary the irreversibility line (blue circles) displays the $(1-t)^2$ dependence characteristic of vortex melting. }
\label{Fig.3}
\end{center}
\end{figure}

\section{upper critical field}

Fig.1 displays typical AC measurements for both $H\|c$ and $H\|ab$ (sample A4). As shown, a well defined specific heat jump is obtained at $T_c$ for $H = 0$ ($\sim 20 \%$ of the total $C_p$) and this peak progressively shifts towards lower temperature as the magnetic field is increased (here up to 28 T). The $H_{c2}$ line has been deduced from the mid-point of the $C_p/T$ anomaly after subtraction of a smooth polynomial background from the raw data. As shown in Fig.2A, the corresponding $H_{c2}$ lines present a very strong downward curvature for $T\rightarrow T_c$ which was not revealed by previous transport measurements (the same behavior is observed in all measured samples, see for instance Fig.2B and Fig.3A for a comparison between samples A4 and B1). Note that a very similar curvature has been reported reported very recently from $C_p$ measurements up to 14T \cite{Sarafin}

Such a curvature is a strong indication for  paramagnetic effects and we have hence fitted the experimental data using a  weak coupling BCS clean limit model including both orbital and Pauli limitations \cite{Brison}. This model only requires two fitting parameters (plus $T_c$) : the initial slope $dH_{c2}/dT|_{T=T_c}$ and the zero temperature Pauli limit $H_p$. The results are shown in Fig.3A for sample  A4 and B1. As shown, very good fits can be obtained in both samples using very similar fitting parameters : $\mu_0dH_{c2}/dT|_{T=T_c} \sim 38 \pm 3$ T/K and $\sim 13 \pm 2$ T/K for $H\|ab$ and $H\|c$ respectively and $\mu_0H_p \sim 45 \pm 2$ T and $\sim 54 \pm 4$ T/K  for $H\|ab$ and $H\|c$, respectively. 

As previously observed in layered systems (see \cite{Ruggiero} and discussion in \cite{Vedeneev}) $H_{c2}^{\|ab}$ is actually very close to a $(1-T/T_c)^{1/2}$ law. Strikingly, this simple behaviour is valid up to $T/T_c \sim 0.99$ in our system (see Fig.3B). Such a dependence can be directly inferred from a Ginzburg-Landau (GL) expansion which leads to  \cite{Vedeneev} :
\begin{equation}
\label{Eq1 }
\left(\frac{H}{H_p}\right)^2+\frac{H}{H_o}=1-\frac{T}{T_c}
\end{equation}
(where $H_o$ is the orbital field) i.e. $H_{c2} \sim H_p(1-t)^{0.5}$ for $H >> H_p^2/H_o$.
A fit to Eq.(1) (solid line in Fig.3B)  leads to $\mu_0H_p^{\|ab} \sim 65$ T and $\mu_0H_p^{\|c}\sim 75$ T, $\mu_0H_o^{\|ab} \sim 650$ T and $\mu_0H_o^{\|c} \sim 170$ T (sample A4) \cite{GLvs0T}. We hence have $\mu_0H_p^2/H_o \sim  6$ T for $H\|ab$, field which is reached for $T/T_c \sim 0.99$. Fe(Se,Te) is thus a rare example of superconductor for which the upper critical field is dominated by paramagnetic effects on almost the totality of the phase diagram (for $H\|ab$). A shown in Fig.3B, a linear dependence is recovered very close to $T_c$ with  $\mu_0dH_o^{\|ab}/dT \sim 45$ T/K and $\mu_0dH_o{\|c}/dT \sim 12$ T/K, in good agreement with a values deduced from the BCS fitting procedure \cite{2Dcrossover}. 

Those extremely high $H_o$ values are related to very small values of the coherence lengths $\xi_{ab} =\Phi_0/2\pi[0.7\times \mu_0H_o] \sim 15 \pm 1 \AA$ and $\xi_c = \xi_{ab}\times(H_o^{\|c}/H_o{\|ab}) \sim 4 \pm 1 \AA$ which confirm the very strong renormalization of the Fermi velocity observed in ARPES measurements \cite{Tamai} (see also theoretical calculations in  \cite{Aichorn}). Indeed, one gets $v_{F,ab}=\pi\Delta\xi_{ab}/\hbar \sim 1.4\times10^4$ m/s ($\Delta$ being the superconducting gap $\sim  2$ meV \cite{Hanaguri,Kato})  i.e. $\hbar v_{F,ab} \sim 0.09$ eV$\AA$ in perfect  agreement with ARPES data which also led to $\hbar v_F\sim 0.09$ eV$\AA$ for the $\alpha_3$ hole pocket centered on the $\Gamma$ point [note that the $H_{c2}$ line will be dominated by the band having the larger critical field i.e. the lower Fermi velocity].  Our measurements do hence confirm the strong correlation effects previously suggested by ARPES measurements \cite{Tamai}. 

An estimate of the paramagnetic field in the weak coupling limit is given by the Clogston-Chandrasekhar formula :	$\mu_0H_p=2\Delta/\sqrt{2}g\mu_B \sim 26$ T in our sample (taking $g=2$) i.e. well below the experimental suggesting that $g \sim 1.0-1.2$. however, it is important to note that $H_p$ may be increased by strong coupling effects \cite{strong coupling} and a fit to the data can be obtained introducing an electron-phonon coupling constant $\lambda \sim 0.6-0.7$ and $g \sim 2$ (still having an anisotropy on the order of 1.2 between the two main crystallographic axis).  Even if it is difficult to conclude on the exact value of $g$, our data clearly indicate a small anisotropy of this coefficient ($\sim 1.2$) supporting the possibility of a crossing of the $H_{c2}$ lines at low temperature. Note that this anisotropy is much lower than the one inferred from transport measurements ($\sim 4$ \cite{Braithwaite}) confirming that the large apparent anisotropy of $g$ deduced from those measurements is an artifact, probably related to the anisotropy of flux dynamics (see discussion on the irreversibiliy line below). The anisotropy of the upper critical field is then strongly temperature dependent rising from $H_{c2}^{ab}/H_{c2}^c \sim H_p^{ab}/H_p^c \sim 0.8$ for $T\rightarrow0$, reflecting the small anisotropy of the g factor, to $H_{c2}^{ab}/H_{c2}^c \sim H_o^{ab}/H_o^c \sim 3.5-4$ close to $T_c$, reflecting the anisotropy of the coherence lengths (see Fig.7). 

 \begin{figure}
\begin{center}
\resizebox{0.48\textwidth}{!}{\includegraphics{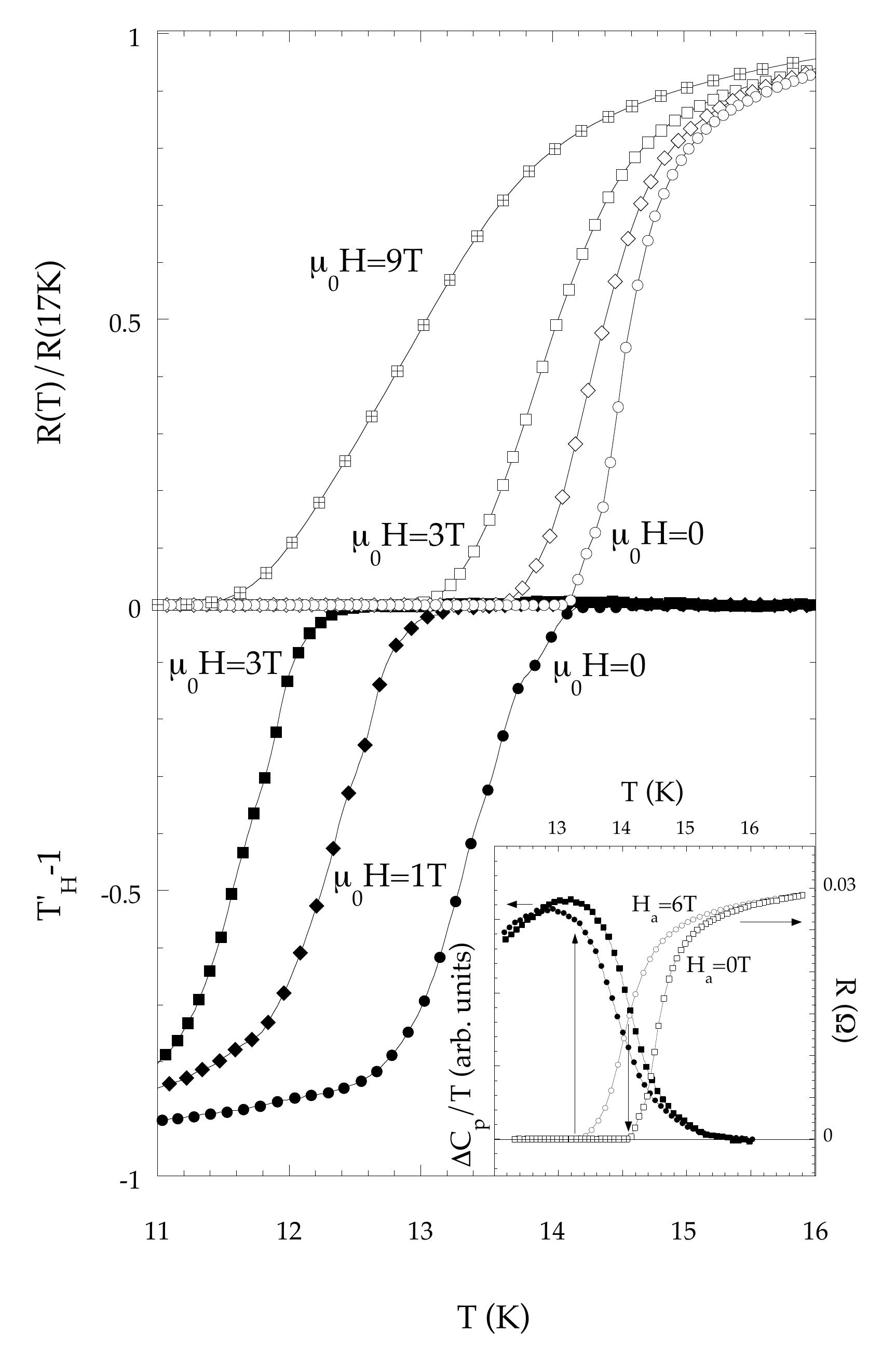}}
\caption{Transport and AC transmitivity measurements as a function of $T$ for the indicated magnetic fields ($H\|c$) in Fe(Se$_{0.5}$,Te$_{0.5}$) single crystals. In the inset : comparison between transport and specific heat data for $\mu_0H=0$ and $6$ T ($\|c$) emphasizing that the midpoint of the specific heat anomaly does not correspond to any characteristic temperature in $R(T)$ for $H\neq0$.}
\label{Fig.4}
\end{center}
\end{figure} 

\section{Irreversibility line}

The small $\xi$ values associated to large $\lambda$ values ($\lambda_{ab}(0) \sim 430$ nm (see below and \cite{Bendele}) lead to strong fluctuation effects hindering any direct determination of $H_{c2}$ from either transport of susceptibility measurements. These fluctuations can be quantified by the Ginzburg number  $G_i = (k_BT_c/\epsilon_0\xi_c)^2/8$ where $\epsilon_0$ ($=(\Phi_0/4\pi\lambda_{ab})^2$) is the line tension of the vortex matter. One hence obtains $\epsilon_0\xi_c \sim 40$ K (as a comparison $\epsilon_0\xi_c \sim 200$K in cuprates) and $G_i \sim 10^{-2}$ which is very similar to the value obtained in YBa$_2$Cu$_3$O$_{7-\delta}$ or NdAsFe(O$_{1-x}$F$_x$) (so called 1111-phase, see \cite{Kacmarcik} and references therein) clearly showing that thermal fluctuations are very strong in this system. 

To emphasize this point,  we have reported in Fig.2, the temperatures corresponding to both $R \rightarrow 0$ and $R/R_N = 0.5$ deduced from transport measurements up to 9T for sample  A4 (see also Fig.4) and even up to 50T for sample B1 (see \cite{Braithwaite}) ($R_N$ being the normal state resistance). As shown, none of those lines present the strong downward curvature obtained in $C_p$ measurements. On the contrary, the $R/R_N=0$ lines vary almost linearly with $T$  with $d\mu_0H/dT \sim 11$ T/K and $\sim 5$ T/K for $H\|ab$ and $H\|c$, respectively in agreement with previous measurements \cite{Lei,Fang}. However, as pointed above, these lines do not correspond to any thermodynamic criterion and discussions of the corresponding lines should hence be taken with great caution. Moreover whereas the midpoint of the specific heat coincides with the $R=0$ temperature for $H=0$ in sample A4, this midpoint rather lies close to the $R/R_N=0.5$ point in sample B1 clearly showing that neither of those two transport criteria can be associated with the $H_{c2}$ line.

Similarly, as previously observed in high temperature cuprates and 1111-pnictides \cite{Kacmarcik}, the onset of the diamagnetic response ($T'_H \rightarrow 0$) also lies well below the the $H_{c2}$ line. (see Fig.2A and Fig.4). Indeed, this onset is related to the irreversibility line above which the system is unable to screen the applied AC field due to the free motion of vortices. This irreversibility line is then expected to lie close to the $R=0$ line. As shown in Fig.4 the onset of diamagnetism actually differs slightly from the onset of resistivity. This difference is much probably related to different voltage-current criteria (the magnetic screening corresponds to much smaller electric fields but requires higher currents) but both lines present the positive curvature characteristic of the onset of irreversible processes.  Note that, as expected for vortex melting (for a review see \cite{Blatter}), the irreversibility line (here defined as the onset of $T'_H$) varies as : $H_{irr} \propto (1-T/T_c)^\alpha$ with $\alpha \sim 2$ (see Fig.3B). A similar curvature has also been reported by Bendele {\it al.} \cite{Bendele} for the irreversibility field deduced from magnetization measurements. 

 \begin{figure}
\begin{center}
\resizebox{0.48\textwidth}{!}{\includegraphics{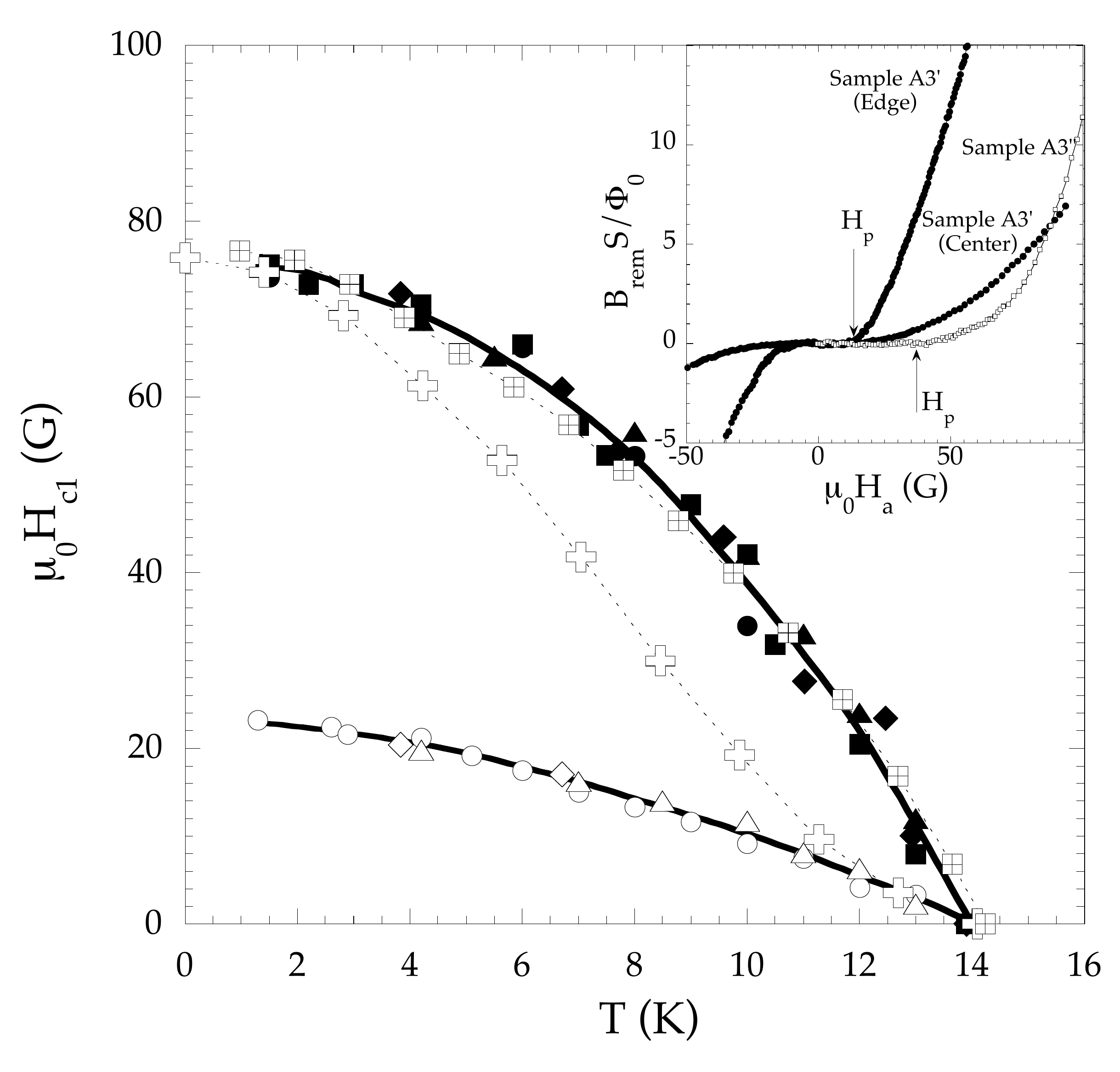}}
\caption{Temperature dependence of the lower critical field ($H_{c1}$) deduced from Hall probe measurements for  $H\|ab$  (open symbols) and $H\|c$ (closed symbols) in sample A1 (squares), A2 (circles), A3 (lozenges) and A3'' (triangles), see Table 1 for sample dimensions. Solid lines are guides to the eyes, crossed squares are muon relaxation data from [23] and open crosses TDO data. In the inset : remanent field $B_{rem}$ as a function of the applied field in sample A3' (center and edge) and A3". }
\label{Fig.5}
\end{center}
\end{figure}

\section{lower critical field}

The first penetration field has been measured on a series of Fe(Se$_{0.5}$Te$_{0.5}$) samples with very different aspect ratios (see Table 1). To avoid spurious effects associated to strong pinning preventing the vortex diffusion to the center of the sample \cite{Okazaki} $H_f$ has also been measured on several locations of the same sample. The inset of Fig.5 displays typical examples on sample A3' (2 positions) and A3". In samples with rectangular cross sections,  flux lines partially penetrate into the sample through the sharp corners even for $H_a < H_f$ but remain "pinned" at the sample equator. The magnetization at $H_a=H_f$ is then larger than $H_{c1}$ and the standard "elliptical" correction for $H_{c1}$ ($=H_f/(1-N)$ where $N$ is the demagnetization factor) can not be used anymore.  Following \cite{Brandt}, in presence of geometrical barriers, $H_{f}$ is related to $H_{c1}$ through : 
\begin{equation} 
H_{c1} \approx \frac{H_f}{tanh(\sqrt{\alpha d /w })} 
\label{Eq.(2)}
\end{equation}
where $\alpha$ varies from 0.36 in strips to 0.67 in disks ($d$ and $w$ being the thickness and width of the sample, respectively. To reduce the uncertainty associated with the $\alpha$ value as well as the $d/w$ ratio in real samples of irregular shape, five different samples with different aspect ratios have been measured (see Table 1). Sample A3' has been cut out of sample A3 and finally A3" out of A3' in order to directly check the influence of the aspect ratio on $H_f$. The corresponding $H_f$ values are reported in the inset of Fig.7 together with the theoretical predictions from Eq.(2) taking $\mu_0H_{c1}^{ab} = 78$ G (the predictions for a an standard "elliptical" correction are also displayed for comparison). 

The lower critical fields ($\mu_0H_{c1}^c$, $\mu_0H_{c1}^{ab}$) are then related to the penetration depth ($\lambda_c$, $\lambda_{ab}$) through : 
\begin{eqnarray}
\mu_0H_{c1}^c=\frac{\Phi_0}{4\pi\lambda_{ab}^2}(Ln(\kappa)+c(\kappa)) \\
\mu_0H_{c1}^{ab}=\frac{\Phi_0}{4\pi\lambda_{ab}\lambda_c}(Ln(\kappa^*)+c(\kappa^*))
\end{eqnarray}
where $\kappa=\lambda_{ab}/\xi_{ab}$, $\kappa* = \lambda_c/\xi_{ab}$ and $c(\kappa)$ is a $\kappa$ dependent function tending towards $\sim 0.5$ for large $\kappa$ values. Taking $\mu_0H_{c2}^c(0) = 0.7\times\mu_0 H_o \sim 130$T, and $H_{c1}^c=78 \pm 5$ G one gets $\lambda_{ab}(0) \sim 430 \pm 50$ nm, which is in fair agreement with  muons relaxation data \cite{Biswas,Bendele}. This very large $\lambda$ value  confirms the general trend previously inferred in iron pnictides  (see for instance \cite{Bendele} and references therein) pointing towards a linear increase of $T_c$ vs $1/\lambda_{ab}^2$ as initially proposed in cuprates by Uemura {\it et al.} \cite{Uemura}. For $H \| ab$,  no correction was introduced (flat samples) and one hence obtains $\mu_0H_{c1}^{ab}=\mu_0H_f^{ab}=23\pm3$ G leading to  $\lambda_c \sim 1600 \pm 200$ nm (taking  $\mu_0H_{c2}^{ab}=0.7\times \mu_0H_o^{ab} \sim 460$ T)

As shown in Fig.5,  $H_{c1}(T)$ clearly shows a saturation at low temperature. As a comparison we have reported on Fig.5 the temperature dependence of the superfluid density deduced from muons relaxation data \cite{Biswas} and $\rho^{TDO}_S(T)$ measurements \cite{Kim}. Both $H_{c1}$ and $\rho^{\mu SR}_S(T)$ curves are similar but do not reproduce the important shoulder at 5 K of the superfluid density. This shoulder has been interpreted as a clear signature of multi-gap superconductivity and as a failure of the clean limit s-wave (including $s\pm$) pairing \cite{Kim}. Our measurements do not support this interpretation.

\begin{figure}
\begin{center}
\resizebox{0.48\textwidth}{!}{\includegraphics{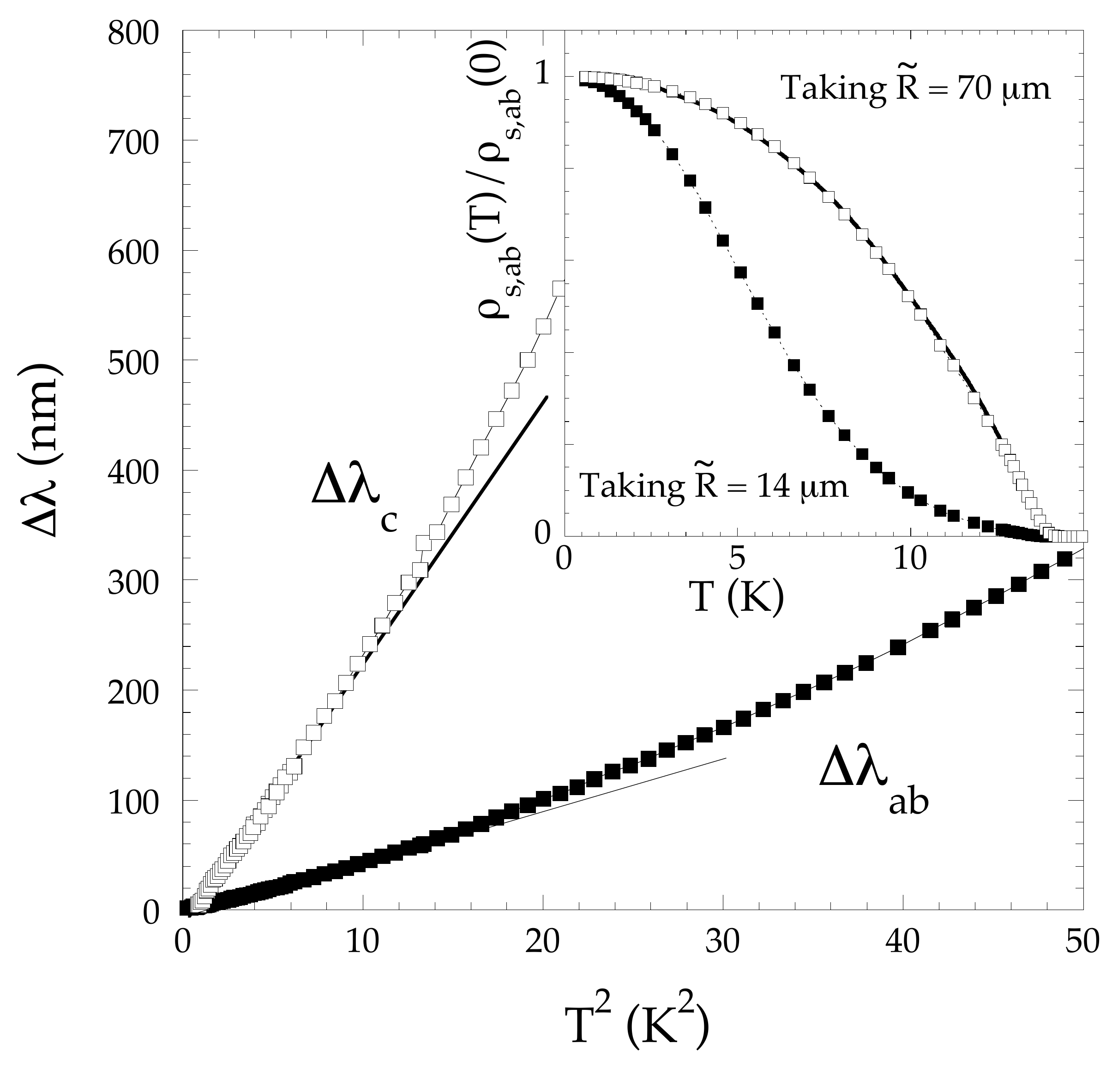}}
\caption{Temperature dependence of $\lambda_{ab}$  (solid symbols) and $\lambda_c$ (open symbols)  deduced from the frequency shift in TDO measurements (sample A3'', Table 1). Inset : temperature dependence of the superfluid density  $\rho^{TDO}_S(T)/\rho^{TDO}_S(0)=1/(1+\Delta\lambda_{ab}(T)/\lambda_{ab}(0))^2$ taking $\lambda_{ab}(0) = 430$ nm and $\tilde{R} = 14 \mu$m (solid symbols) (i.e.  following [22] , see corresponding  $\Delta\lambda_{ab}(T)$ values on  the main panel) or  $\tilde{R} =70 \mu$m (open symbols). The average $H_{c1}(T)/H_{c1}(0)$ curve (see Fig.5) is displayed as the thick solid line.}
\label{Fig.6}
\end{center}
\end{figure}

In order to shed light on this discrepancy, we have performed TDO measurements on each of the samples of Table 1. As described in sec.II, $\lambda_c$ and $\lambda_{ab}$ were deduced from the effective penetration depth $\widetilde{\lambda}$ measured for both $H\|ab$ and $H\|c$. As shown in Fig.6 (sample A3''), both $\Delta\lambda_{ab}$ and $\Delta\lambda_c$ are proportional to $T^n$ with n close to 2, in good agreement with previous measurements for $H\|c$ \cite{Kim} (the same temperature dependence has been obtained for all samples). The TDO data then require the introduction of the value of $\lambda_{ab}(0)$ to convert the $\Delta\lambda(T)$ data into $\rho^{TDO}_S(T)/\rho^{TDO}_S(0)=1/(1+\Delta\lambda_{ab}(T)/\lambda_{ab}(0))^2$. Introducing $\lambda_{ab}(0)\sim 430nm$ and taking $\tilde{R} \ \sim 14 \mu$m (from \cite{Prozorov}), $\rho^{TDO}_S(T)$ shows a change of curvature around 5K, very similar to the one previously reported in \cite{Kim} (see inset of Fig.6). A similar discrepancy has already been observed in MgCNi$_3$ and interpreted as a reduction of the critical temperature at the surface of the sample due to a modification of the carbon stoechiometry \cite{Diener}. However, such an explanation is not expected to hold here as single crystals were extracted mechanically from the bulk. 

It is important to note that the temperature dependence of the superfluid density is very sensitive to the absolute value of $\Delta\lambda$ and, although very similar to the one reported by Kim {\it et al.} \cite{Kim}, the amplitude of  $\Delta\lambda_{ab}/T^2 \sim 40 \AA$/K$^2$ observed in our samples is much larger than the one reported recently by Sarafin {\it et al.} \cite{Sarafin} ($\sim 10 \AA$/K$^2$). Similar discrepancies in the absolute amplitude of $\Delta \lambda$ have also been reported in other pnictides \cite{Hashimoto} and have been attributed to complications from rough edges which may lead to an overestimation of $\Delta \lambda$. Dividing the absolute $\Delta \lambda_{ab}$ by a factor $\sim 5$  (i.e. taking $\tilde{R}=70$ $\mu$m for $H\|c$ instead of $14$ $\mu$m)  actually leads to a very good agreement between TDO and $H_{c1}$ data (see Fig.6) hence indicating that this value has probably been overestimated due to an underestimation of the effective dimension $\tilde{R}$ in presence of rough edges.

Very similar temperature dependences of $H_{c1}$ were obtained in both directions (see Fig.5) leading to a (almost) temperature independent anisotropy of $H_{c1}$ : $\Gamma_{H_{c1}} \sim 3.4 \pm 0.5$ and hence $\Gamma_\lambda = \lambda_c/\lambda_{ab}= [H_{c1}^c/H_{c1}^{ab}]\times[(Ln(\kappa^*)+c(\kappa))/(Ln(\kappa^*)+c(\kappa))]\sim \Gamma_{H_{c1}}\times1.2 \sim 4.1 \pm 0.8$ (see Fig. 7). This value is hence very close to the one obtained for $H_{c2}$ close to $T_c$ as $\Gamma_{H_{c2}}(T\rightarrow T_c) \sim \Gamma_{H_o}=\xi_{ab}/\xi_c$ (see Fig.7). Similarly, very similar temperature dependences have been obtained for $\Delta\lambda_c$ and $\Delta\lambda_{ab}$ (with $\Delta\lambda_c \sim 5 \times \Delta\lambda_{ab}$ up to $T \sim T_c$) again suggesting a weak temperature dependence of this anisotropy. Finally, this value is also close to the one obtained for the irreversibility field deduced from the onset of diamagnetic screening.

\begin{figure}
\begin{center}
\resizebox{0.48\textwidth}{!}{\includegraphics{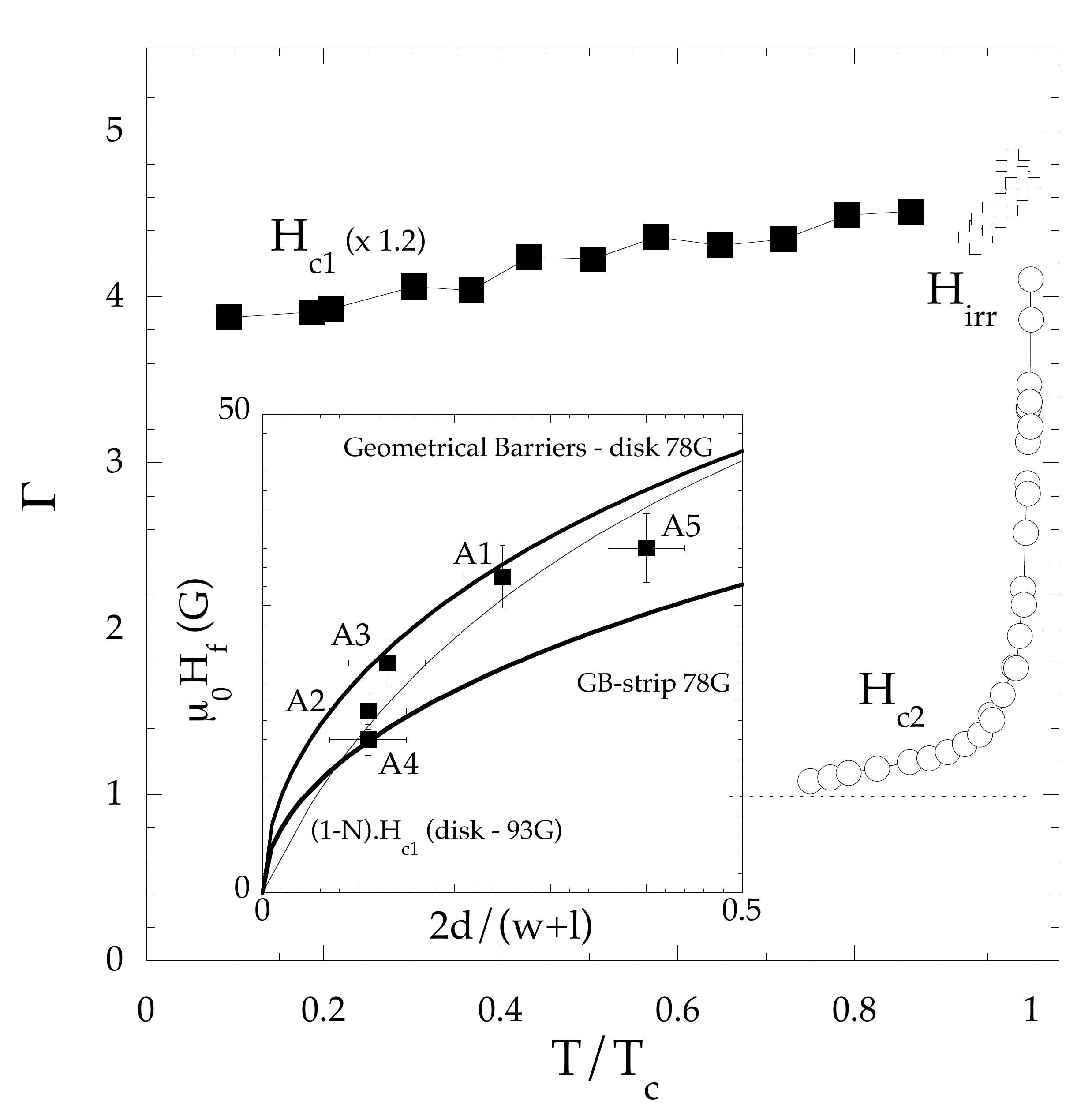}}
\caption{Temperature dependence of the anisotropy of the upper (open circles), and lower (solid squares) critical fields and irreversibility line (onset of diamagnetic response, open crosses). The $\Gamma_{H_{c1}}$ data have been multiplied by a factor 1.2 to display $\Gamma_\lambda = 1.2 \times \Gamma_{H_{c1}}$. In the inset : first penetration field $H_f$ as a function of the aspect ratio :  $2d/(w+l)$. Thick solid lines are theoretical predictions in presence of geometrical barriers using see Eq.(2) for disks (upper line) and strips (lower line); the thin line correspond to a standard  "elliptical" correction (no surface barrier).}
\label{Fig.7}
\end{center}
\end{figure}

\section{final discussion}
 
The value of the normal state Sommerfeld coefficient ($\gamma_N$) in Fe(Se,Te) compounds remains debated as values ranging from $\sim 23$ mJ/molK$^2$ \cite{Tsurkan} to $\sim 39$ mJ/molK$^2$ \cite{Sales} have been obtained. For non superconducting samples, it has even been shown recently \cite{Liu2} that $\gamma_N$ rises rapidly for $x < 0.1$ reaching $\sim 55$ mJ/molK$^2$ for $0.1 \leq x \leq 0.3$.  Even though our maximum field (28 T) is too low to fully destroy superconductivity down to 0 K  hence hindering any precise determination of $\gamma_N$, it is worth noting that a $\gamma_N$ value on the order of $\sim 39$ mJ/molK$^2$ is incompatible with the entropy conservation rule in our sample. A reasonable fit to the data (solid line in Fig.1C) assuming that $C_p/T = \gamma_N+\beta T^2 + \delta T^4$ for $20>T>12$ K and $\mu_0H= 28$ T leads to $\gamma_N=23\pm 3$ mJ/molK$^2$ in good agreement with the value obtained by by Tsurkan {\it et al.} \cite{Tsurkan}. This $\gamma_N$ value is also in fair agreement with the one deduced from ARPES measurements ($\sim 30$ mJ/molK$^2$ \cite{Tamai}). Similarly, the Debye temperature  ($\Theta_D \sim 143$ K) is in reasonable agreement with the one previously reported in both Fe(Se$_{0.67}$Te$_{0.23}$) ($\Theta_D \sim 174$ K \cite{Sales}) and Fe$_{1.05}$Te ($\Theta_D \sim 141$ K \cite{Chen}).  
 
The electronic contribution to the specific heat ($C_e/T=C_p/T-\beta T^2-\delta T^4$) is then displayed in the inset of Fig.1 together with the theoretical prediction for a single gap BCS superconductor in the weak coupling limit (i.e. taking $2\Delta/kT_c \sim 3.5$, thin solid line). As shown, this standard behavior largely overestimates the experimental data at low temperature suggesting the presence of a much larger gap. A reasonable agreement to the data is obtained assuming that $2\Delta/kT_c \sim 5$ (dotted line). However, even though some indication for the presence of a large gap were obtained by fitting either $\mu SR$ \cite{Biswas} or optical conductivity \cite{Homes} data, the corresponding gap value ($\sim 3$ meV) is much larger than the value obtained by spectroscopy ($\sim 1.8-2$ meV \cite{Hanaguri,Kato}). Moreover those former measurements also suggest the presence of a much smaller gap which is not present in our specific heat meaurements.
 
Some evidence for nodes (or for deep gap minima) in Fe(Se$_{0.5}Te_{0.5}$) has been suggested by four fold oscillations in the low temperature specific heat for $H\|c$ \cite{Zeng}. However, despite the high resolution of our AC technique and the very good quality of our samples (the specific heat jump at $T_c$ is slightly larger than in \cite{Zeng}) we did not observe these oscillations in our samples (i.e $\Delta C_p(\theta)/C_p < 10^{-3}$). Nodes are also expected to show up in the field dependence of the Sommerfeld coefficient ($\gamma(H)$) which is then expected to vary as $H^\alpha$ with $\alpha < 1$ ($\alpha = 0.5$ for the so-called Volovik effect for d-wave pairing with line nodes whereas $\alpha \sim 1$ for classical single gap BCS systems). We have hence extrapolated the $C_e(H)/T$ data to zero using either a BCS formula (see discussion above, $C_e/T-\gamma(H) \propto exp(-\Delta(H)/kT)$ in our temperature range) or a phenomenological second order polynomial fit. Both procedure led to a {\it concave} curvature for $\gamma(H)$ with $\alpha \sim 1.5\pm0.3$ for $H\|c$ and $\alpha \sim 2.2\pm0.6$ for $H\|ab$. This concave behavior can be attributed to the effect of Pauli paramagnetism on the vortex cores \cite{Adachi} (see \cite{Aoki,Nishizaki} for experimental data in heavy fermions) hence clearly supporting the importance of these effects in Fe(Se$_{0.5}$Te$_{0.5}$). 

Finally note that it has been suggested that $\Delta C_p/T_c$ could be proportional to $T_c^2$  in iron pnictides \cite{Budko,Paglione} due to strong pair breaking effects \cite{Kogan} with $\Delta C_p/T_c^3 \sim 0.06$ mJ/molK$^4$. One hence would expect an anomaly $\Delta C_p/T_c \sim 12$ mJ/molK$^2$ at $T_c$ in our system which is clearly lower than the experimental value $ \sim 40 \pm 5$ mJ/molK$^2$. Similarly, it has been suggested that the initial slope of the $H_{c2}$ line could scale as $\mu_0dH_{c2}^{c}/dT \sim 0.2\times T_c$ (T/K) but, again, this scaling does not hold in our sample for which $\mu_0dH_{c2}^{c}/dT \sim 12$ T/K.  Finally note that the temperature dependence of the superfluid density (see discussion above) supports the $\Delta\lambda_{ab}/T^2 \sim 10 \AA$/K$^2$ value obtained by Serafin {\it et al.} \cite{Sarafin} which is also much smaller than the one suggested from the scaling of \cite{Gordon} : $\Delta\lambda_{ab}/T^2 \sim 8.8 \times 10^4/T_c^3 \sim 32  \AA$/K$^2$. 

\section{conclusion}

In summary, 

(i)  Precise determinations of the $H_{c2}$ lines from $C_p$ measurements led to a very strong downward curvature, similar to that observed in layered systems. 

(ii) The temperature dependence of the upper critical field and the field dependence of the Sommerfeld coefficient both indicate that $H_{c2}$ is limited by strong paramagnetic effects with $\mu_0H_p \sim 45\pm2$ T and $\sim 54\pm4$ T for $H\|ab$ and $H\|c$, respectively.

(iii) The very small value of the coherence length $\xi_{ab}(0) \sim 15 \AA$ confirms the strong renormalisation of the effective mass (compared to DMFT calculations) previously observed in ARPES measurements \cite{Tamai} and associated strong electron correlation effects. $\gamma_N$ is estimated to $\sim 23\pm 3$ mJ/molK$^2$ in fair agreement with the ARPES value.

(iv) The anisotropy of the orbital critical field is estimated to be on the order of $4$ hence leading to a $\xi_c(0)$ value smaller than the c lattice parameter.

(v) Neither the temperature dependence of $\lambda$ nor that of the electronic contribution to the specific heat follow the weak coupling BCS model (an BCS dependence with $\Delta/kT_c \sim 5$ remains possible) but no evidence for nodes in the gap is obtained from the field dependence of the Sommerfeld coefficient. We did not observe the fourfold oscillations of the low temperature specific heat previously obtained by Zeng {\it et al.} \cite{Zeng}.

(vi) The amplitude of the specific heat jump $\Delta C_P/T_c \sim 40 \pm 5$ mJ/molK$^2$ is much larger than that previously observed in Fe(Se,Te) and does not follow the $\Delta C_p/T_c^3$ inferred in iron pnictides. Similarly neither the slope of the $H_{c2}$ line nor the absolute value of $\Delta\lambda(T)$ obey the scaling laws previously proposed for iron pnictides \cite{Kogan,Gordon}.

(vii) $\lambda_{ab}(0) = 430 \pm 50$ nm and $\lambda_c(0) = 1600 \pm 200$ nm, confirming the very small superfluid density previously observed in iron pnictides. The corresponding anisotropy is almost temperature independent with $\Gamma_\lambda \sim \Gamma_{H_{c2}}(T\rightarrow T_c)=\gamma_\xi$.

(viii) These large $\lambda$ values associated to small $\xi$ values lead to a very small condensation energy $\epsilon_0\xi_c \sim 40$K and hence to large fluctuation effects hindering any determination of $H_{c2}$ from either transport or susceptibility measurements. A detailed analysis of the influence of these fluctutations on the specific heat anomaly will be presented elsewhere.

(ix) The strong upward curvature of the irreversibility line (defined as the onset of diamagnetic screening) : $H_{irr} \propto (1-T/T_c)^2$ strongly suggests the existence of a vortex liquid in this system. 

This work has been supported by the French National Research Agency,
Grant ANR-09ÐBlanc-0211 ÕSupraTetraferÕ and ANR 'DELICE', and by the Euromagnet II grant via the EU co tract RII-CT-2004-506239. TK is most obliged to V. Mosser of ITRON, Montrouge,
and M.Konczykowski from the Laboratoire des Solides Irradi\'es, Palaiseau for the development of the Hall sensors used in this study. We thank J-P.Brison for the software used to fit the $H_{c2}$ data.


\begin{references}    

\bibitem{Kamihara} Y. Kamihara, T. Watanabe, M. Hirano, and H. Hosono, J. Am. Chem. Soc. {\bf 130}, 
3296 (2008); X. H. Chen, T. Wu, G. Wu, R. H. Liu, H. Chen, and D. F. Fang, Nature London  {\bf 453}, 761 (2008).

\bibitem{Hsu} F. C. Hsu, J. Y. Luo, K. W. Yeh, T. K. Chen, T. W. Huang, P. M. Wu, Y. C. Lee, Y. L. Huang, Y. Y. Chu, D. C. Yan, and M. K. Wu, Proc. Natl. Acad. Sci. U.S.A, {\bf 105}, 14262 (2008).

\bibitem{Garbarino} Garbarino {\it et al.} Euro. Phys. Lett. {\bf 86}, 27001(2009); Medvedev {\it et al.} Nature Mat. {\bf 8}, 630 (2009); Margadonna {\it et al.} Phys. Rev. B {\bf 80}, 064506 (2009); Mizuguchi {\it et al.} Applied Phys. Lett. {\bf  93}, 152505 (2008), Braithewaite {\it et al.} J. Phys.: Condens. Matter {\bf 21}, 232202 (2009).

\bibitem{Sales} B. C. Sales, A. S. Sefat, M. A. McGuire, R. Y. Jin, D. Mandrus, and Y. Mozharivskyj, Phys. Rev. B {\bf 79}, 094521 (2009).

\bibitem{Yeh} K.-W. Yeh, T. W. Huang, Y. L. Huang, T. K. Chen, F. C. Hsu,P. M. Wu, Y. C. Lee, Y. Y. Chu, C. L. Chen, J. Y. Luo, D. C. Yan, and M. K. Wu, Europhys. Lett. {\bf 84}, 37002 (2008); M. H. Fang, H. M. Pham, B. Qian, T. J. Liu, E. K. Vehstedt, Y. Liu, L. Spinu, and Z. Q. Mao, Phys. Rev. B, {\bf 78}, 224503 (2008).

\bibitem{Fe2} Assuming that the additional iron atoms siting on Fe2 sites between the Se/Te atoms do not constitute a charge reservoir.

\bibitem{Bao} Wei Bao, Y. Qiu, Q. Huang, M. A. Green, P. Zajdel, M. R. Fitzsimmons, M. Zhernenkov, S. Chang, M. Fang, B. Qian, E. K. Vehstedt, Jinhu Yang, H. M. Pham, L. Spinu, and Z. Q. Mao, Phys. Rev. Lett. {\bf 102}, 247001 (2009).

\bibitem{Qiu} Y.Qiu,W.Bao, Y. Zhao, C. Broholm, V. Stanev, Z. Tesanovic, Y. C. Gasparovic, S. Chang, Jin Hu, Bin Qian, M.Fang, and Z. Mao, Phys. Rev. Lett. {\bf 103}, 067008 (2009); M. D. Lumsden, A. D. Christianson, E. A. Goremychkin, S. E. Nagler, H. A. Mook, M. B. Stone, D. L. Abernathy, T. Guidi, G. J. MacDougall, C. de la Cruz, A. S. Sefat, M. A. McGuire, B. C. Sales and D. Mandrus, Nature Physics, {\bf 6}, 182 (2010).

\bibitem{Martinelli} A. Martinelli, A. Palenzona, M. Tropeano, C. Ferdeghini, M. Putti, M. R. Cimberle, T. D. Nguyen,M. Affronte, and C. Ritter, Phys. Rev. B, {\bf 81}, 094115 (2010).

\bibitem{Tamai} A. Tamai, A.Y. Ganin, E. Rozbicki, J. Bacsa, W. Meevasana, P. D. C. King, M. Caffio, R. Schaub, S. Margadonna, K. Prassides, M. J. Rosseinsky, and F. Baumberger, Phys. Rev. Lett. {\bf 104}, 097002 (2010).

\bibitem{Lei} H.Lei,  R.Hu, E. S. Choi, J. B. Warren, and C. Petrovic, Phys. Rev. B, {\bf 81}, 094518, (2010); T.Kida, T.Matsunaga, M. Hagiwara, Y.Mizuguchi,  Y.Takanoe, and K.Kindo, J. Phys. Soc. Jpn {\bf 78}, 113701 (2009).

\bibitem{Fang} M. H. Fang, J. H. Yang, F. F. Balakirev, Y. Kohama, J. Singleton, B. Qian, Z. Q. Mao, H. D. Wang, H. Q. Yuan, hys. Rev. B {\bf 81}, 020509 (2010)

\bibitem{Braithwaite} D.Braithwaite, G.Lapertot, W.Knafo, I.Sheikin, J. Phys. Soc. Jpn {\bf 79}, 053703 (2010).

\bibitem{Sarafin} A. Serafin, A. I. Coldea, A.Y. Ganin, M.J. Rosseinsky, K. Prassides, D. Vignolles, and A. Carrington, Phys. Rev. B, {\bf 82}, 104514 (2010).

\bibitem{Pribulova} Pribulova, T. Klein, J. Kacmarcik, C. Marcenat, M. Konczykowski, S. L. BudÕko, M. Tillman, and P. C. Canfield Phys. Rev. B {\bf 79}, 020508 (2009).

\bibitem{Yadav} C.S.Yadav, P.L.Pauloose, New Journal of Physics {\bf 11}, 103046 (2009).

\bibitem{Bendele} M. Bendele, S. Weyeneth, R. Puzniak, A. Maisuradze, E. Pomjakushina, K. Conder,
V. Pomjakushin, H. Luetkens, S. Katrych, A. Wisniewski, R. Khasanov, and H. Keller, Phys. Rev. B, {\bf 81}, 224520 (2010).

\bibitem{Kim} H. Kim, C. Martin, R. T. Gordon, M. A. Tanatar, J. Hu, B. Qian, Z. Q. Mao, Rongwei Hu, C. Petrovic, N. Salovich, R. Giannetta, and R. Prozorov, Phys. Rev. B, {\bf 81}, 180503(R), (2010).

\bibitem{Budko} S. L. BudÕko, Ni Ni, and P. C. Canfield, Phys. Rev. B {\bf 79}, 220516(R) (2009).

\bibitem{Paglione} J.Paglione and R.L.Greene, Nat. Phys. {\bf 6}, 645 (2010).

\bibitem{Liu} T. J. Liu, X. Ke, B. Qian, J. Hu, D. Fobes, E. K. Vehstedt, H. Pham, J. H. Yang, M. H. Fang, L. Spinu, P. Schiffer, Y. Liu, and Z. Q. Mao, Phys. Rev. B {\bf 80}, 174509 (2009).

\bibitem{Prozorov} R. Prozorov and R. W. Giannetta Superconductor Science and Technology \textbf{19}, R41 (2006).

 \bibitem{Brison} J. P. Brison, N. Keller, A. Verniere, P. Lejay, L. Schmidt, A. Buzdin, J. Flouquet, S. R. Julian, and G. G. Lonzarich: Physica C 250 (1995) 128.

\bibitem{Ruggiero} S.T.Ruggiero, T.W.Barbee and M.R.Beasley, Phys. Rev. Lett. {\bf 49}, 1299 (1980); C.Uher, J.L.Cohn and I.K.Schuller, Phys. Rev. B, {\bf 34}, 4906 (1986).

\bibitem{Vedeneev} S. I. Vedeneev, C. Proust, V. P. Mineev, M. Nardone, and G. L. J. A. Rikken, Phys. Rev. B, {\bf 73}, 014528 (2006). 

\bibitem{GLvs0T} note that the the zero temperature extrapolations of the fields deduced from GL expansions always overestimate the experimental values.

\bibitem{2Dcrossover} Note that we could expect to observe a dimensional crossover at a temperature $T^*$ for which $\xi_c(T^*) \sim c/\sqrt{2}$ with $H_{c2}^{ab} = \sqrt{3}\Phi_0/\pi d\xi_{ab} \propto (1-T/T_c)^{0.5}$ for $T<T^*$ (taking $\xi_{ab} =\xi_{ab}(0)\times(1-T/T_c)^{-0.5}$). This crossover is however hindered by the Pauli field $H_p << H_{c2}^{2D} \sim 2600\times(1-T/T_c)^{0.5}$. 

\bibitem{Aichorn} M.Aichhorn, S.Biermann, T.Miyake, A.Georges, and M.Imada, Phys. Rev. B, {\bf 82}, 064504 (2010).

\bibitem{Hanaguri} T. Hanaguri, S. Niitaka, K. Kuroki, H. Takagi, Science {\bf 328}, 474 (2010).

\bibitem{Kato} T.Kato, Y.Mizuguchi, H. Nakamura, T.Machida, H.Sakata, and Y.Takano, Phys. Rev. B, {\bf 80}, 180507(R) (2009).

\bibitem{strong coupling} $H_p$ is expected to be enhanced by a factor $\eta(1+\lambda_{e-ph})^\epsilon$ in the strong coupling limit where $\lambda_{e-ph}$ is the electron-phonon coupling constant and $\epsilon$ is an exponent ranging from 0.5 (T. P. Orlando {\it et al.}Phys. Rev. B {\bf 19}, 4545 (1979) to 1 (M. Schossmann and J. P. Carbotte, Phys. Rev. B {\bf 39}, 4210 (1989). 

\bibitem{Kacmarcik} J.Kacmarcik, C.Marcenat, T.Klein, Z.Pribulova, M.Konczykowski, S. L. Budko, M. Tillman, N.Ni and P. C. Canfield, Phys. Rev. B,  {\bf 80}, 014515 (2009).

\bibitem{Blatter}  G. Blatter, M. V. FeigelÕman, V. B. Geskenbein, A. I. Larkin, and V. M. Vinokur, Rev. Mod. Phys. {\bf 66}, 1125 (1994).

\bibitem{Okazaki} R. Okazaki, M. Konczykowski, C. J. van der Beek, T. Kato, K. Hashimoto, M. Shimozawa, H. Shishido, M. Yamashita, M. Ishikado, H. Kito, A. Iyo, H. Eisaki, S. Shamoto, T. Shibauchi,1 and Y. Matsuda, Phys. Rev. B, {\bf 79}, 064520 (2009).

\bibitem{Brandt} E.H. Brandt Phys. Rev. B {\bf 59}, 3369 (1999).

\bibitem{Biswas} P. K. Biswas, G. Balakrishnan, D. M. Paul, C. V. Tomy,,  M. R. Lees,1 and A. D. Hillier, Phys. Rev. B, {\bf 81}, 092510 (2010).

\bibitem{Uemura} J. Uemura, G. M. Luke, B. J. Sternlieb, J. H. Brewer, J. F.Carolan, W. N. Hardy, R. Kadono, J. R. Kempton, R. F. Kiefl, S.R. Kreitzman, P. Mulhern, T. M. Riseman, D. Ll, Williams, B.
X. Yang, S. Uchida, H. Takagi, J. Gopalakrishnan, A.W. Sleight,M. A. Subramanian, C. L. Chien, M. Z. Cieplak, G. Xiao, V. Y.Lee, B. W. Statt, C. E. Stronach, W. J. Kossler, and X. H. Yu, Phys. Rev. Lett. {\bf 62}, 2317 (1989).

\bibitem{Diener} P. Diener, P. Rodi\`ere, T. Klein, C. Marcenat, J. Kacmarcik, Z. Pribulova, D. J. Jang, H. S. Lee, H. G. Lee, and S. I. Lee, Phys. Rev. B \textbf{79}, 220508 (2009)

\bibitem{Hashimoto} K. Hashimoto, A. Serafin, S. Tonegawa, R. Katsumata,
R. Okazaki, T. Saito, H. Fukazawa, Y. Kohori, K. Kihou,
C. H. Lee, A. Iyo, H. Eisaki, H. Ikeda, Y. Matsuda, A. Carrington, and T. Shibauchi, Phys. Rev. B {\bf 82}, 014526 (2010).

\bibitem{Tsurkan} V. Tsurkan, J. Deisenhofer, A. Gunther, C. Kant, H.-A. Krug von Nidda,F. Schrettle, and A. Loidl, cond-mat arXiv: 1006.4453v1.

\bibitem{Liu2} T. J. Liu, J. Hu, B. Qian, D. Fobes, Z. Q. Mao, W. Bao, M. Reehuis, S. A. J. Kimber, K. Prokes, S. Matas, D. N. Argyriou, A. Hiess, A. Rotaru, H. Pham, L. Spinu, Y. Qiu, V. Thampy, A. T. Savici8,, J. A. Rodriguez and C. Broholm, Nat. Mat. {\bf 9}, 716 (2010).

\bibitem{Chen} G. F. Chen, Z. G. Chen, J. Dong, W. Z. Hu, G. Li, X. D. Zhang, P. Zheng, J. L. Luo, and N. L. Wang, Phys. Rev. B, {\bf 79}, 140509(R) (2009). 

\bibitem{Homes} C.C. Homes, A. Akrap, J.S. Wen, Z.J. Xu, Z.W. Lin, Q. Li, and G.D. Gu, Phys. Rev. B 81, 180508 (2010).

\bibitem{gamma0} note that the zero field curve extrapolates towards a very small but non zero value which can be attributed to the presence of a small fraction of parasitic phase.

\bibitem{Zeng} B. Zeng, G. Mu, H. Q. Luo, T. Xiang, H. Yang, L. Shan,  C. Ren, I. I. Mazin, P. C. Dai  and H.-H. Wen, cond-mat arXiv:1007.3597v1.

\bibitem{Adachi}  H.Adachi, M.Ichiode, K.Machidaz, J.Phys. Soc. Jpn {\bf 74} 2181 (2005); M.Ichioda and K.Machida, Phys. Rev. B, {\bf 76} 064502 (2007).

\bibitem{Aoki} H. Aoki, T. Sakakibara, H. Shishido, R. Settai, Y. Onuki, P. Miranovi, and K. Machida, J. Phys. Condens. Matter {\bf 16}, L13 (2004);  K. Deguchi, S. Yonezawa, S. Nakatsuji, Z. Fisk, and Y. Maeno, J.Magn. Magn. Mater. {\bf 310}, 587 (2007).

\bibitem{Nishizaki} S. Nishizaki, Y. Maeno and Z. Mao, Journal of the Phys. Soc. of Jpn. {\bf 69}, 572 (2000).

\bibitem{Kogan} V.G.Kogan, Phys. Rev. B {\bf 80}, 214532 (2009).

\bibitem{Gordon} R. T. Gordon, H. Kim, M. A. Tanatar, R. Prozorov, and V. G. Kogan, Phys. Rev. B, {\bf 81}, 180501(R) (2010).

 \end{references}
\end{document}